\documentclass[pra,twocolumn,superscriptaddress]{revtex4}
\usepackage{amsmath,amssymb,graphicx,xcolor,bm,soul}

\begin{document}

\title{Optically induced transparency in bosonic cascade lasers}

\author{T. C. H. Liew}
\affiliation{Division of Physics and Applied Physics, School of Physical and Mathematical Sciences, Nanyang Technological University, 21 Nanyang Link, Singapore 637371}

\author{A. V. Kavokin}
\affiliation{School of Physics and Astronomy, University of Southampton, Southampton SO17 1BJ, United Kingdom}
\affiliation{CNR-SPIN, Viale del Politecnico 1, I-00133, Rome, Italy}

\begin{abstract}
Bosonic cascade lasers are terahertz (THz) lasers based on stimulated radiative transitions between bosonic condensates of excitons or exciton-polaritons confined in a trap.
We study the interaction of an incoming THz pulse resonant in frequency with the transitions between neighboring energy levels of the cascade. We show that at certain optical pump
conditions the cascade becomes transparent to the incident pulse: it neither absorbs nor amplifies it, in the mean field approximation. The populations of intermediate levels of the bosonic cascade change as the THz pulse passes, nevertheless. In comparison, a fermionic cascade laser does not reveal any of these properties.
\end{abstract}

\date{\today}

\maketitle

The concept of bosonic cascade lasers has been introduced a few years ago with the objective of generating THz frequency radiation in a compact semiconductor system~\cite{Liew2013}. The bosonic cascade is defined as a series of bosonic energy levels with equal THz range spacing in energy. A boson excited in the highest level can undergo a series of transitions down the cascade, generating multiple THz frequency photons in analogy to fermionic quantum cascade lasers~\cite{Kazarinov1971,Faist1994}, which were also developed in the THz regime~\cite{Kohler2002,Williams2007,Liu2013,Belkin2015}. The critical difference of a bosonic cascade is that bosonic final state stimulation enhances the scattering rates such that even in the limit of weak spontaneous scattering rate, particles can reach the ground level of the cascade. A variety of theoretical considerations of bosonic cascades have since been considered, including the quantum statistics of the cascade levels~\cite{Liew2016} and the interplay of double bosonic stimulation coming from a THz cavity and the bosonic particles themselves~\cite{Kaliteevski2014}. Physical implementations of bosonic cascades can be based on excitons or exciton-polaritons in parabolic traps~\cite{Tosi2012} and are presently under experimental development~\cite{Tzimis2015,Trifonov2016}.

In this Letter, we discuss the electromagnetically induced transparency (EIT) of bosonic cascades. EIT is currently studied in a large variety of systems including diluted atomic gases~\cite{Morigi2000,Haller2015}, solid solutions, electromechanical~\cite{Teufel2011}, optomechanical systems~\cite{SafaviNaeini2011}, etc. EIT is especially promising for the slowing and storing of light~\cite{Bajcsy2003}. EIT in Bose-Einstein condensates has been extensively discussed as well, see e.g.~\cite{Hau2008}. In this context, we find that bosonic cascades offer an interesting peculiarity linked with the interplay of stimulated absorption and emission of light in the cascade that leads to the transparency. This is in contrast with the interference nature of EIT in the most part of the aforementioned works.

We define the transparency of a system at a given frequency as the property of allowing electromagnetic radiation to pass at that frequency. This definition can still be applied to a system that is already generating radiation at the chosen frequency. In such case, the definition only means that there should be no change to the generation rate when the system is illuminated, or if some radiation is absorbed, an equal amount should be re-emitted such that there is no net change to the generation-absorption rate.

\section{Operation Scheme} 
\begin{figure}[h!]
\includegraphics[width=\columnwidth]{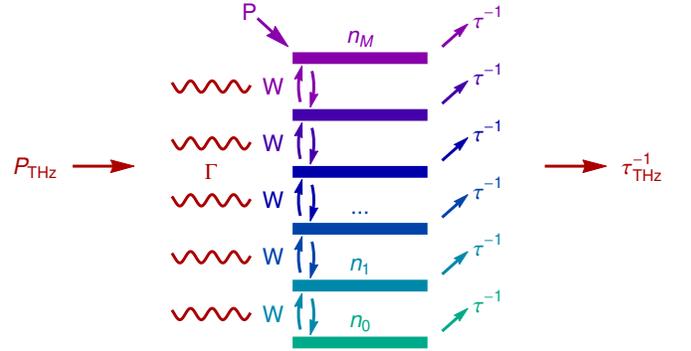}
\caption{Generic scheme of a bosonic cascade laser. A series of equally separated energy levels with population $n_k$ are coupled via THz frequency radiative transitions with spontaneous scattering rate $W$. The highest level in the cascade is resonantly driven, at rate $P$, and all cascade levels dissipate energy, at rate $\tau^{-1}$. An incident THz field $P_\mathrm{THz}$ is considered, which leaves the system at a rate $\tau_\mathrm{THz}$. We denote $\Gamma$ as the net terahertz generation rate.} \label{fig:scheme}
\end{figure}

We consider a cascade of $M$ bosonic levels, with populations $n_k$, coupled to a THz mode, with population $n_\mathrm{THz}$, as illustrated in Fig.~\ref{fig:scheme}.

Such a system has been described previously by the system of semiclassical rate equations for the cascade level populations $n_k$~\cite{Liew2013}:
\begin{align}
\frac{dn_0}{dt}=-\frac{n_0}{\tau}&+W\left[n_1\left(n_0+1\right)\left(n_\mathrm{THz}+1\right)\right.\notag\\
&\hspace{5mm}\left.-n_0\left(n_1+1\right)n_\mathrm{THz}\right]\label{eq:dn0dt}
\end{align}
\begin{align}
\frac{dn_k}{dt}=-\frac{n_k}{\tau}&+W\left[n_{k+1}\left(n_k+1\right)\left(n_\mathrm{THz}+1\right)\right.\notag\\
&\hspace{5mm}\left.-n_k\left(n_{k+1}+1\right)n_\mathrm{THz}\right]\notag\\
&+W\left[n_{k-1}\left(n_k+1\right)n_\mathrm{THz}\right.\notag\\
&\hspace{5mm}\left.-n_k\left(n_{k-1}+1\right)\left(n_\mathrm{THz}+1\right)\right]\notag\\
&\hspace{15mm}\forall\hspace{5mm}0<k<M
\end{align}
\begin{align}
\frac{dn_m}{dt}=P-\frac{n_M}{\tau}&+W\left[n_{M-1}\left(n_M+1\right)n_\mathrm{THz}\right.\notag\\
&\hspace{5mm}\left.-n_M\left(n_{M-1}+1\right)\left(n_\mathrm{THz}+1\right)\right]\label{eq:dnMdt}
\end{align}
Here $W$ is the spontaneous transition rate between neighbouring levels, $\tau$ is the lifetime of particles in each level, and $P$ is the pumping rate. For simplicity, we assume that $W$ and $\tau$ are independent of the level index. We also neglect higher order scattering processes (e.g., parametric scattering processes~\cite{Diederichs2006}), which were considered in detail in Ref.~\cite{Liew2013}.

To model an incident THz field, we assume the existence of a THz mode, with intensity $n_\mathrm{THz}$, that overlaps with the bosonic cascade modes. This THz mode is driven externally at a rate $P_\mathrm{THz}$, experiences gain $\Gamma$ from relaxation processes in the cascade, and leaves the system with a lifetime $\tau_\mathrm{THz}$:
\begin{equation}
\frac{dn_\mathrm{THz}}{dt}=-\frac{n_\mathrm{THz}}{\tau_\mathrm{THz}}+\Gamma+P_\mathrm{THz}\\\label{eq:dnTHzdt}
\end{equation}
In the absence of any confinement of the THz mode (that is, in the absence of any THz cavity), the lifetime can be estimated from the time it takes a THz photon to cross the cascade. For example, considering a $3\mu$m system size, a THz photon would take approximately $0.01ps$ to cross the system going at the speed of light (assuming a refractive index of the system material equal to $1$ at THz frequency). Even though the lifetime of a THz photon in the system is very short, it is important to describe the THz field in the system by a dynamical equation to allow for the possibility of its depletion.

We define $\Gamma$ as the net THz generation-absorption rate, which is given by summing all the stimulated THz emission processes and subtracting the absorption processes:
\begin{align}
\Gamma&=W\sum_{k=0}^{M-1}\left[n_{k+1}\left(n_k+1\right)n_\mathrm{THz}-n_k\left(n_{k+1}+1\right)n_\mathrm{THz}\right]\notag\\
&=W(n_M-n_0)n_\mathrm{THz}\label{eq:Gamma}
\end{align}
Note that only THz photons generated by processes stimulated by $n_\mathrm{THz}$ are included in $\Gamma$. Spontaneously generated THz photons are accounted for in the equations for the evolution of the cascade levels, however they would be emitted in all directions while it is implicit that $n_\mathrm{THz}$ represents a composition of THz photons traveling in a particular direction through the system. Our objective is to study the change in the net THz generation-absorption rate, that is, $\Gamma$, when the system is subjected to a THz field.

\section{Steady-state solution} 
\begin{figure}[h!]
\includegraphics[width=\columnwidth]{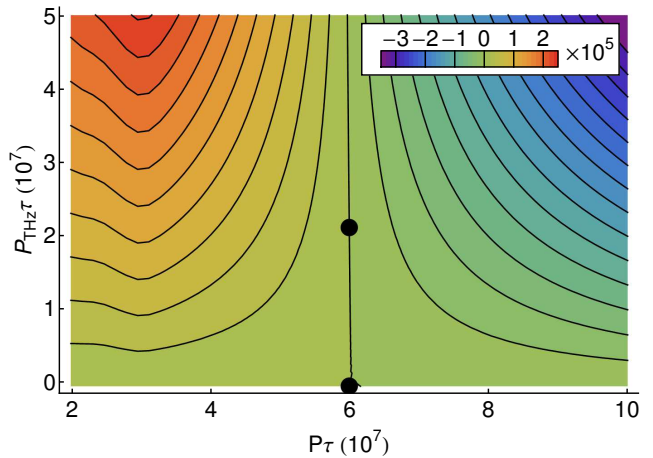}
\caption{Contour plot of the dependence of $\Gamma$ on $P$ and $P_\mathrm{THz}$. The black spots indicate an example of values where $\Gamma$ is unchanged from switching $P_\mathrm{THz}$ on and off. Parameters: $W\tau=8.3\times10^{-7}$ (this value is taken from calculations of transition matrix elements of excitons in parabolic quantum wells~\cite{Liew2013}), $\tau_\mathrm{THz}=10^{-3}\tau$.} \label{fig:GammaColour}
\end{figure}

Equations~\ref{eq:dn0dt}-\ref{eq:dnTHzdt} are readily solved for the steady-state of the system. Figure~\ref{fig:GammaColour} shows the variation of the net THz generation-absorption rate, $\Gamma$, as a function of $P$ and $P_\mathrm{THz}$, for typical parameters. It can be seen that for small cascade pumping, the application of a THz field increases the THz generation rate, while for large cascade pumping, the application of a THz field reduces the THz generation rate.

This effect can be interpreted by considering that the THz field tends to favour equal populations of all cascade levels, as it can either enhance relaxation of particles in the cascade (through stimulating THz emission) or enhance excitation of particles in the cascade (through THz absorption). Since the cascade pump, $P$, is applied to the highest level in the cascade it favours population of higher levels at weak intensity. Application of the THz field then favours the relaxation of particles from these high levels resulting in THz emission. On the other hand, a strong cascade pump favours relaxation to the lower levels of the cascade, due to strong stimulated THz emission by cascade particles. In this regime, the THz pump favours excitation of the cascade levels, corresponding to absorption of THz photons.

The vertical contour in Fig.~\ref{fig:GammaColour} that divides the regions of positive and negative $\Gamma$ is most important for our purposes. The existence of this contour makes $\Gamma$ almost independent on $P_\mathrm{THz}$ at a selected value of $P$. Although it does not show on the scale of the plot, the contour is very slightly curved, however, it is still possible to find specific values of $P$ and $P_\mathrm{THz}$ for which $\Gamma$ is exactly zero in the presence and absence of $P_\mathrm{THz}$ (these values are marked as black spots in Fig.~\ref{fig:GammaColour}). Under these conditions, we would expect a continuous THz field to pass through the system with no change to its intensity.

\section{Dynamics} 
\begin{figure}[h!]
\includegraphics[width=\columnwidth]{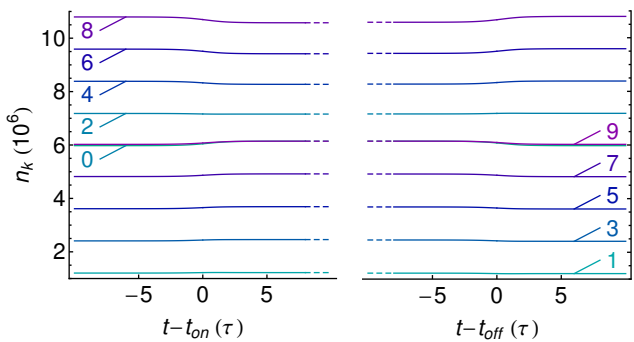}
\caption{Evolution of the bosonic cascade level populations during a THz pulse with square profile (numbers in labels refer to the index of the cascade level). The pulse is characterized by a switch on time $t=t_\mathrm{on}$ (in the left-hand plot) and a switch off time $t=t_\mathrm{off}$ (in the right-hand plot). The parameters were the same as in Fig.~\ref{fig:GammaColour}, taking $P\tau=6\times10^7$. The square pulse was given sigmoid edges (of width $\tau$) and a height $P_\mathrm{THz}\tau=2.14\times10^7$.} \label{fig:cascadebosonic}
\end{figure}

The dynamics of bosonic cascades subject to pump pulses can be calculated by propagating Eqs.~\ref{eq:dn0dt}-\ref{eq:dnTHzdt} numerically in time. Fig.~\ref{fig:cascadebosonic} considers an initial steady state of the cascade and then the application of a THz pulse. We do not specialize to a pulse of any specific duration, but assume that it is longer than the timescales set by $\tau$ and $\tau_\mathrm{THz}$. In the time intermediate between switch on and switch off of the square pulse, the system is effectively in a steady state. This allows us to make general statements about square pulses with different durations. Given that polaritons may have lifetimes on the picosecond timescale, it is implied that we consider pulses of duration on the order of tens of picoseconds or longer. The intensity of the square pulse and the cascade pump intensity is taken to correspond to the black spots in Fig.~\ref{fig:GammaColour}.

Although the relative changes are not very large, they are clearly noticeable. Consequently, the THz pulse is in principle detectable for all times within the pulse duration.
\begin{figure}[h!]
\includegraphics[width=\columnwidth]{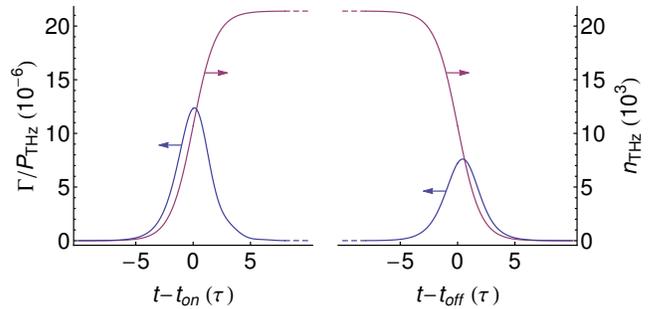}
\caption{Evolution of $\Gamma$ and $n_\mathrm{THz}$ during a THz pulse with square profile. Parameters were the same as in Fig.~\ref{fig:cascadebosonic}} \label{fig:GammaEvolution}
\end{figure}
At the same time, Fig.~\ref{fig:GammaEvolution} shows the evolution of $\Gamma$ when the THz pulse is switched on and off. There are transient changes in $\Gamma$ upon switch on and off of the THz pulse, however, they are extremely small, being on the order of $10^{-6}$ compared to the THz field pump rate. Indeed this can be expected, as in an adiabatic regime one follows the almost vertical contour in Fig.~\ref{fig:GammaColour}. In practice, since the contour is not perfectly vertical, small changes in $\Gamma$ occur when moving between the two steady states, and working in a non-adiabatic regime equally allows for keeping the changes in $\Gamma$ small. After the transient effects have died out, $\Gamma$ is exactly zero after switch on or off of the THz pulse, corresponding to a perfect transparency of the system to the THz field.

For completeness, Fig.~\ref{fig:GammaEvolution} shows also the evolution of the THz mode population, $n_\mathrm{THz}$. Since $\Gamma$ is always very small, the THz mode population essentially follows the sigmoid switch on and off of $P_\mathrm{THz}$. While we have considered square shaped pulses (with sigmoid edges), for which the physics of the system can be interpreted as making transitions between the steady states of Fig.~\ref{fig:GammaColour}, other pulse shapes could also be considered. More abruptly changing pulse shapes give a larger transient contribution to $\Gamma$, and for continuously varying pulse shapes (e.g., Gaussian) there will always be in general a non-zero value of $\Gamma$ at all times. In principle, with Gaussian shaped pulses it is possible to arrange for the time integrated value of $\Gamma$ to be zero, through careful choice of $P$ and $P_\mathrm{THz}$. In such case the time-integrated intensity of a THz pulse would be preserved, however, due to non-zero instantaneous values of $\Gamma$ the THz pulse would become reshaped in time.

\section{Comparison with Fermionic Cascades} It is instructive to compare the physics of bosonic cascades with fermionic cascades, which obey a similar set of rate equations to Eqs.~\ref{eq:dn0dt}-\ref{eq:dnTHzdt}, but with the stimulated terms due to cascade occupation removed:
\begin{align}
\frac{dn_0}{dt}=-\frac{n_0}{\tau}&+W\left[n_1\left(n_\mathrm{THz}+1\right)-n_0n_\mathrm{THz}\right]\\
\frac{dn_k}{dt}=-\frac{n_k}{\tau}&+W\left[n_{k+1}\left(n_\mathrm{THz}+1\right)-n_kn_\mathrm{THz}\right]\notag\\
&+W\left[n_{k-1}n_\mathrm{THz}-n_k\left(n_\mathrm{THz}+1\right)\right]\notag\\
&\hspace{15mm}\forall\hspace{5mm}0<k<M\\
\frac{dn_m}{dt}=P-\frac{n_M}{\tau}&+W\left[n_{M-1}n_\mathrm{THz}-n_M\left(n_\mathrm{THz}+1\right)\right]
\end{align}
\begin{align}
\frac{dn_\mathrm{THz}}{dt}&=-\frac{n_\mathrm{THz}}{\tau_\mathrm{THz}}+\Gamma+P_\mathrm{THz}
\end{align}
where we now define $\Gamma$ according to:
\begin{align}
\Gamma&=W\sum_{k=0}^{M-1}\left[n_{k+1}n_\mathrm{THz}-n_kn_\mathrm{THz}\right]\notag\\
&=W(n_M-n_0)n_\mathrm{THz}\label{eq:GammaFermionic}
\end{align}

A direct comparison of fermionic and bosonic cascades using the same parameters is not particularly intuitive. Bosonic cascades are designed to function in the limit $W\tau\ll1$, where they make use of the bosonic stimulation of scattering processes to allow efficient relaxation through all levels of the cascade. Fermionic cascades can not function in the limit $W\tau\ll1$; taking the typical value $W\tau=8.3\times10^-7$~\cite{Liew2013} for bosonic cascades and substituting into the fermionic case will leave a system with no significant population of all but the highest level.

\begin{figure}[h!]
\includegraphics[width=\columnwidth]{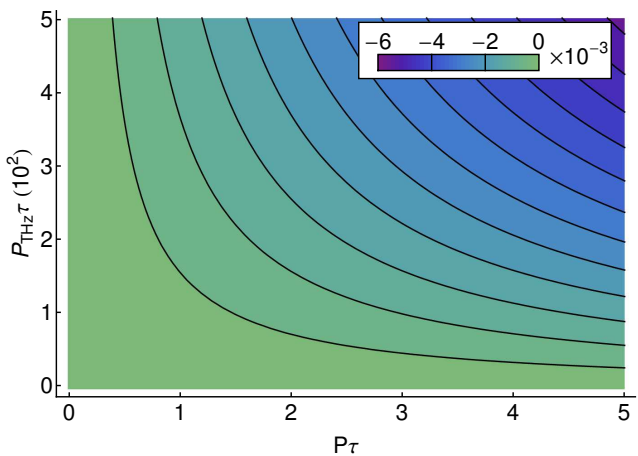}
\caption{Dependence of $\Gamma$ on $P$ and $P_\mathrm{THz}$ for a fermionic cascade. Parameters: $W\tau=10$, $\tau_\mathrm{THz}=1.688\times10^{-4}\tau$.} \label{fig:GammaColourFermions}
\end{figure}

We can nevertheless consider the possibility of a fermionic cascade achieving transparency, in the limit $W\tau>1$ using a similar analysis to the bosonic case. Figure~\ref{fig:GammaColourFermions} shows a contour plot of $\Gamma$ on $P$ and $P_\mathrm{THz}$.
Unlike the bosonic case, the contours $\Gamma(P_\mathrm{THz})=\mathrm{constant}$ increase monotonically in $P_\mathrm{THz}$ and it is impossible to find parameters such that $\Gamma$ is unchanged when turning $P_\mathrm{THz}$ on and off.

\section{Conclusion} The transparency in bosonic cascades has a different physical origin compared to the well-known EIT. It does not rely on the negative interference of pump and probe fields. Rather, it stems from the exact compensation of the absorption and stimulated emission of radiation, and can be described in the mean-field approximation neglecting the phase of light and interference effects. On the quantum optical level, the bosonic cascades are not transparent, strictly speaking. However, at certain conditions, the mean number of photons in a light pulse going through the cascade would remain unchanged. This peculiar property makes bosonic cascades promising for applications as non-destructive photo-detectors. THz light passes going through bosonic cascades would leave traces in the instantaneous redistributions of mean populations of the cascade levels. None of these effects can be found in fermionic cascades.

AK acknowledges the support from the EPSRC Programme grant on Hybrid Polaritonics. TL acknowledges support from the Ministry of Education (Singapore, grant 2015-T2-1-055).



\begin{thebibliography}{1}

\bibitem{Liew2013}
T C H Liew, M M Glazov, K V Kavokin, I A Shelykh, M A Kaliteevski, \& A V Kavokin, Phys. Rev. Lett., {\bf 110}, 047402 (2013),

\bibitem{Kazarinov1971}
R F Kazarinov \& R A Suris, Sov. Phys. Semiconductors, {\bf 5}, 707 (1971).

\bibitem{Faist1994}
J Faist, F Capasso, D L Sivco, C Sirtori, A L Hutchinson, \& A Y Cho, Science, {\bf 264}, 553 (1994).

\bibitem{Kohler2002}
R Kohler, A Tredicucci, F Beltram, H E Beere, E H Lineld, A Giles Davies, D A Ritchie, R C Lotti, \& F
Rossi, Nature, 417, 156 (2002).

\bibitem{Williams2007}
B S Williams, Nature Photon., {\bf 1}, 517 (2007).

\bibitem{Liu2013}
J Liu, J Chen, T Wang, Y Li, F Liu, L Li, L Wang, \& Z Wang, Solid State Comm., {\bf 81}, 68 (2013).

\bibitem{Belkin2015}
M A Belkin \& F Capasso, Phys. Scr., {\bf 90}, 118002 (2015).

\bibitem{Liew2016}
T C H Liew, Y G Rubo, A S Sheremet, S De Liberato, I A Shelykh, F P Laussy, \& A V Kavokin, New J. Phys., {\bf 18}, 023041 (2016).

\bibitem{Kaliteevski2014}
M A Kaliteevski, K A Ivanov, G Pozina, \& A J Gallant, Sci. Rep., {\bf 4}, 5444 (2014).

\bibitem{Tosi2012}
G Tosi,	G Christmann, N G Berloff, P Tsotsis, T Gao, Z Hatzopoulos,	P G Savvidis, \& J J Baumberg, Nature Phys., {\bf 8}, 190 (2012).	

\bibitem{Tzimis2015}
A Tzimis, A V Trifonov, G Christmann, S I Tsintzos, Z Hatzopoulos, I V Ignatiev, A V Kavokin, \& P G Savvidis, Appl. Phys. Lett., {\bf 107}, 101101 (2015).

\bibitem{Trifonov2016}
A V Trifonov, E D Cherotchenko, J L Carthy, I V Ignatiev, A Tzimis, S Tsintzos, Z Hatzopoulos, P G Savvidis, \& A V Kavokin, Phys. Rev. B, {\bf 93}, 125304 (2016).

\bibitem{Morigi2000} 
G Morigi, J\"urgen Eschner, \& C H Keitel, Phys. Rev. Lett., {\bf 85}, 4458 (2000).

\bibitem{Haller2015}
E Haller, J Hudson, A Kelly, D A Cotta, B Peaudecerf, G D Bruce, \& S Kuhr, Nature Phys., {\bf 11}, 738 (2015).

\bibitem{Teufel2011}
J D Teufel, D Li, M S Allman, K Cicak, A J Sirois, J D Whittaker \& R W Simmonds, Nature, {\bf 471} 204 (2011).

\bibitem{SafaviNaeini2011}
A H Safavi-Naeini, T P Mayer Alegre, J Chan, M Eichenfield,	M Winger, Q Lin, J T Hill, D E Chang, \& O Painter, Nature, {\bf 472} 69 (2011).

\bibitem{Bajcsy2003}
M Bajcsy, A S Zibrov, \& M D Lukin, Nature, {\bf 426}, 638 (2003).

\bibitem{Hau2008}
L V Hau, Nature Photon., {\bf 2}, 451 (2008).

\bibitem{Diederichs2006}
C Diederichs, J Tignon, G Dasbach, C Ciuti, A Lemaitre, J Bloch, P Roussignol, \& C Delalande, Nature, {\bf	440}, 904 (2006).


\end{thebibliography}
\end{document}